\documentstyle[12pt]{article}
\begin{document}

\newcommand{\beq}{\begin{equation}}
\newcommand{\eeq}{\end{equation}}
\newcommand{\beqa}{\begin{eqnarray}}
\newcommand{\eeqa}{\end{eqnarray}}
\renewcommand{\d}{{\rm d}}
\newcommand{\e}{{\rm e}}
\newcommand{\s}{{\rm s}}
\renewcommand{\P}{{\rm P}}
\newcommand{\k}{{\bf k}}
\newcommand{\y}{{\bf y}}
\newcommand{\vsl}{v \hspace{-.17cm} / }
\renewcommand{\b}{\tilde{b}}
\renewcommand{\le}{\stackrel{{\textstyle <}}{_ {\displaystyle \sim}}}
\renewcommand{\ge}{\stackrel{{\textstyle >}}{_ {\displaystyle \sim}}}

\begin{titlepage}

\begin{flushright}
ITP--SB--93--83
\end{flushright}

\vspace{1cm}

\begin{center}
{\bf  PROTON-PROTON NEAR-FORWARD  HARD ELASTIC SCATTERING }
\vspace{1.5cm}

Michael G. Sotiropoulos  and George Sterman\\
\vspace{1cm}
{\em Institute for Theoretical Physics, State University of New York, \\
       Stony Brook, N.Y. 11794-3840, U.S.A.}

\vspace{.5cm}

\today

\vspace{1cm}

\end{center}

\begin{abstract}

 We calculate the leading twist contribution to near-forward
 proton-proton (and proton-antiproton)
 elastic scattering with large momentum transfer, in the
 multiple scattering (Landshoff) mechanism.
 The amplitude in the near-forward region is dominated by
 singlet exchange for all three valence quark-quark scatterings.
 We assume the existence of a hard singlet quark-quark
 amplitude, which we estimate to be ${\cal O}(\alpha_s^2/t)$.
 For a three-quark state whose transverse size is less than
 $1/\Lambda_{\rm QCD}$, Sudakov resummation accounts for both approximate
 $\d \sigma^{pp}/ \d t \sim t^{-8}$ at moderate $t$, and
 $\d \sigma^{pp}/ \d t \sim t^{-10}$ at larger $t$.
 The transition from approximate
 $t^{-8}$ to $t^{-10}$
 behavior is strongly correlated with the transverse size of the valence
 three-quark state in the proton.

\end{abstract}

\begin{flushleft}
--------------------------------------------------- \\
{\footnotesize
e-mail: michael@max.physics.sunysb.edu \\
e-mail: sterman@max.physics.sunysb.edu \\}
\end{flushleft}

\end{titlepage}
\baselineskip=18pt

\section{Introduction}

 Perturbative QCD (pQCD) offers a formalism in which to study high energy
 exclusive hadronic processes \cite{dimctg,pQCDexcl,BL}.
 Factorization theorems make possible the separation of
 long distance effects,
 describing the formation of color singlet hadronic bound states,
 from the short distance interactions of
 quarks and gluons. Once long distance interactions
 have been modeled by hadronic wave functions,
 the remaining hard process can be calculated perturbatively.
 At leading power, the relevant partons are
 the {\em valence} quarks that carry the momenta of the incoming and
 outgoing hadrons.

 Fixed-angle hadron-hadron elastic scattering has been discussed
 in this formalism,
 for both a single hard scattering \cite{dimctg,pQCDexcl,BL} and for
 multiple, independent (`Landshoff'), scatterings of the
 valence quarks \cite{Land,DonnLand}.
 To calculate the amplitude, however, it is necessary to
 incorporate systematically the
  effect of soft gluon exchange, as
 described in refs.\ \cite{BS,Botts}. This extended
 factorization can also be useful in the treatment of
 hadronic form factors \cite{LiS,Li} and proton-antiproton annihilation
 \cite{Hyer}.
 The resulting suppression
 of spatially separated partonic scatterings
 accounts for approximate ``quark counting" power behavior \cite{dimctg}
 for fixed angle hadron-hadron elastic scattering.
  In this paper, we will show that the improved perturbative formalism,
 including the transverse structure of the proton, can naturally
 account for the approximate $t^{-8}$ behavior of
 $\d \sigma^{pp}/ \d t$ at small angles. In addition, we suggest a
 relatively sharp transition to dimensional counting behavior,
 $\d \sigma^{pp}/ \d t \sim t^{-10}$, as $t$ increases.
 The position of this transition is directly related to the
 transverse size of the quark valence state of the proton.

 Large angle scattering ($\theta ={\cal O}(\pi/2)$) involves two scales,
 the momentum transfer  $-t \sim s$, and the scale of hadronic masses.
 In contrast,
 near-forward scattering ($\theta \rightarrow 0$) is a three scale
 problem, involving the center of mass energy, $\sqrt{s}$,
 the momentum transfer $-t$ and the scale of hadronic masses.
 Correspondingly, the near-forward amplitude at each order has both
  ``Sudakov" double logarithms in momentum transfer and
 ``Regge" logarithms of the energy \cite{Reg1,BFKL}.
 In a recent paper, we have studied the relationship between these two
 phenomena \cite{SS1}

In this paper, we use the results of Refs.\ \cite{SS1} and
\cite{BS} to investigate the interplay between
 soft gluon corrections and the tranverse structure of the proton
 in a range of  momentum  transfers,
 $3\, {\rm GeV}^2 \le -t \le 40 \,{\rm GeV}^2$ .
 For the description of the hadronic
 bound state we use quark distribution amplitudes derived from QCD
 sum rules \cite {CZnuclphys,COZ,KS,GS}.
 To leading power in $\alpha_s$, the three valence quarks from each proton
 scatter independently.
 Of course, at the lower end of the $t$-range considered
 here not all individual q-q scatterings can be really hard.
 Nevertheless, the Sudakov-improved pQCD approach
 remains well defined, and  affords
 a model for the amplitude over the entire range.

 In the $s \rightarrow \infty$, $t$ fixed, limit q-q
 elastic scattering is
 dominated by  singlet exchange in the $t$-channel.
 The suppression of octet exchange may be described as
  gluon reggeization \cite{BFKL}.
 In \cite{SS1} we
 have studied this process as the $\theta \rightarrow 0$ limit of fixed
 angle q-q scattering.
 There, we rederived the reggeization of
 octet exchange from the energy-dependence of its
 Sudakov suppression, while the corresponding factor for the singlet
 is $s$-independent in the leading logarithmic approximation.
 Of course, even with singlet exchange, the elastic scattering of
 colored objects is IR divergent.
 In \cite{SS1} we have suggested ways of separating dimensionally
 regularized
 IR singular contributions from the hard part of
 singlet exchange. The renormalization group can be used to
 resum the leading and non-leading $\ln(t)$ dependence of
 the q-q amplitude, employing one-loop anomalous dimensions
 \cite {BS,Lipquasel,Korch}.

 In the following, we shall treat hard
 singlet exchange as an $\alpha_s^2$ process \cite{as2pomeron}
 only (denoted $H^m_\s(Q_m)$ below),
 where $m=1,2,3$ labels the hard q-q scattering of momentum transfer
 $Q_m$.
 Once the q-q scatterings are embedded in
 the full hadronic amplitude, IR divergences cancel,
 due to the color singlet nature of the external hadrons. The
 IR cutoff (denoted  $1/\b$ below),
 although arbitrary when q-q scattering is considered
 in isolation \cite{SS1}, has a physical meaning
 in the case of proton-proton scattering,
 as the transverse separation between independent
 hard scatterings.

 The expiclit form of the hard singlet to lowest order,
 $\alpha_s^2$, is ambiguous, because the singlet amplitude
 is already IR divergent to this order. As pointed out
 in ref. \cite{SS1}, different IR subtractions procedures yield
 different expressions for $H_\s^m$. The IR subtractions that define
 the hard singlet to lowest order can be fixed only by their
 effect on the amplitude at higher orders in $\alpha_s$.
 Without facing this difficult question here,
 we estimate the lowest order hard singlet in section 3 below.

 Because, as noted above,
 the transverse separation between hard scatterings
 plays the role of an IR cutoff, the transverse structure of
 the proton valence state wave function is an essential
 feature of our calculation.  The perturbative (small ${\tilde b}$)
 behavior of the wave function is available from \cite{BS},
 but for the range of $t$
 considered here we will find it necessary to include as
 well a ``nonperturbative" or ``intrinsic" transverse
 structure as well as the familiar light cone
 wave functions.  The importance of transverse structure has
 been stressed in
 \cite{Banff,Dziem,JacKroll}.

 Our discussion is organized as follows.
 In section 2 we describe in some detail
  the factorized form of the p-p amplitude
 in the multiple scattering scenario \cite{BS}
 for fixed scattering angle $\theta$,
 stressing the roles of
 hadronic wave functions and flavor flows.
In section 3,  we take this amplitude
 to the forward region,
 considering the contribution from singlet exchange in all three
 q-q scatterings. In section 4, we calculate the lowest order
 three-singlet amplitude. This calculation
 determines the combination of the quark distribution amplitudes
 that appears in the p-p amplitude.  Leading and
 non-leading $\ln(t)$ corrections from all orders are then included through
  soft gluon resummation
  factors. In section 5, we define
 the kinematic cut-offs and the choice of scales that turn the closed form
 of the p-p amplitude into a numerically stable four dimensional integral,
 which we compute for a range of momentum transfers.
 We examine the rate of convergence of the amplitude in impact parameter
 space and its sensitivity to scale choices, and study
 the effect of the transverse structure of the valence
 three-quark state.
 We summarize our results in the final section.

\section{The proton-proton amplitude in the multiple scattering scenario}

 We consider the leading twist contribution of the valence three-quark
 Fock state for each proton to the
 proton-proton elastic scattering amplitude.
 The p-p amplitude, ${\cal A}$, is shown in  fig.\ 1 for
 multiple scattering \cite{Land}.
 The  valence quark momenta are parameterized in the CM frame as
\beq
k_{i}^{\mu}=\sqrt{\frac{s}{2}} x_{i} v^{\mu}_{i}
            + \kappa_{i} v^{\prime \mu}_{i} +
        \k^{\mu}_{i} \, ,
\label{kparam}
\eeq
 where $i$ is the proton label, the $v_{i}$ are lightlike vectors
 along the proton directions, with
 $P_{i}=\sqrt{s/2} \: v_{i}$, and
 the $v^{\prime}_{i}$ are lightlike vectors
 in the directions opposite to  $v_{i}$ ($v_{i}\cdot v^{\prime}_{i}=1$).
 $\k_i^\mu \equiv k_{i \perp}^{\mu}$ denotes the components of $k_i$ in
 the plane perpendicular to the spacelike component of $v_i$.
 Using the notation $v_{ij}= v_{i}\cdot v_{j}$,
 we parameterize the invariants of the scattering as
\beq
t=-sv_{13}=-sv_{24}, \hspace{1cm} u=-sv_{14}=-sv_{23} \, .
\label{tu}
\eeq

 Since each proton contains two $u$-quarks
 and one $d$-quark in the valence state,
 there is more than one flavor-distinguishable way of connecting
 the quarks to the three hard scatterings.  We
 label the inequivalent flavor flows as follows. We call $H^3$ the hard
 scattering blob in which the $d$-quark from proton 2 participates, and $H^2$
 the hard scattering in which the $d$-quark from proton 1 participates,
 if $H^2 \neq H^3$.
 Then we obtain the three inequivalent flavor flows shown in fig.\ 1,
 labeled by $f=1,2,3$. Notice that for $f=1$, $H^1$ and  $H^2$
 are flavor indistinguishable.

 For identical quarks, both $t$ and $u$ gluon exchange channels are
 available.
 For $f=2$ the gluonic exchange channels in
 the u-d hard scatterings ($H^2$ and $H^3$)
 are fixed to be $t$,  and for  $f=3$ they are fixed to be $u$.
 Table 1 summarizes the channel configurations
 of the hard scatterings for each flavor flow.

 \subsection{The factorized amplitude}

 According to ref.~\cite{BS},
the leading twist elastic p-p amplitude
 may be written in the following factorized form,
\beqa
{\cal A}(s,t,h_{i}) &=& (-i)\frac{(2\pi)^{8}}{\sin^{2}\theta}
   \int_{0}^{1} \d\xi_1 \d\xi_2
   \int \prod_{i=1}^{4} \d \kappa_{i} \d^{2} \k_{i}
   \d \hat{\kappa_{i}} \d^{2} \hat{\k}_{i}
   \int \d\ell \d\hat{\ell}
    \nonumber \\
&\ & \times \sum_{f=1}^{3} \sum_{ \stackrel{\alpha_{i},}{a_{i},}
    \stackrel{ \beta_{i},} {b_{i},}\stackrel{\gamma_{i}}{c_{i}} }
    Y^{(f)}_{ \alpha_{i} \beta_{i}\gamma_{i}}(k_{i},\hat{k}_{i};
    P_{i},h_{i}) \nonumber \\
&\ & \times U_{\{ a_{i} b_{i} c_{i}\} }(\ell,\hat{\ell})
    \delta(\eta \cdot \sum_{i} \k_{i}+\ell)
       \delta(\eta \cdot \sum_{i} \hat{\k}_{i}+\hat{\ell})
    \nonumber \\
&\ & \times H^1_{\{ \alpha_{i},a_{i}\}}(\xi_1 P_i)
    H^2_{\{ \beta_{i},b_{i} \}}(\xi_2 P_i)
    H^3_{\{ \gamma_{i},c_{i} \}}((1-\xi_1-\xi_2) P_i) + R \, ,
    \nonumber \\
\label{factorizedA}
\eeqa
 where $\alpha_{i},a_{i}$ are Dirac and ${\rm SU}(N_{c})$ indices
 respectively,
 $h_{i}$ are proton helicities, $\theta$ is the scattering angle in the CM
 frame, $\eta$ is the spacelike unit vector perpendicular to the scattering
 plane and $R$ is a remainder suppressed by a power of  $\sqrt{s}$.
 The $Y$'s in eq.\ (\ref{factorizedA}) represent the valence structure
 of the external protons, the $H^m$'s
 represent the three hard scatterings
 and $U$ summarizes the effects of soft gluons, in terms of
 two transverse momentum components
 normal to the scattering plane, $\ell$ and $\hat{\ell}$.
 From the kinematics of q-q scattering
 when all the valence quarks are on shell,
 $x_i = \xi_1 \geq 0$, $\hat{x}_i = \xi_2 \geq 0$
 in eq.~(\ref{kparam}) for every $i=1,2,3,4$.
 The antisymmetric color structure of each wave function has
 been separated from the $Y$'s and absorbed into the
 soft-gluon function $U$.  $U$ has remaining
 free color indices that are summed against those of the
 hard scatterings, $H^m$,
 as in Ref.~\cite{SS1}.

 The hard scatterings depend only on the large components of the
 quark momenta.  For this reason we may integrate
 the wave functions over the
 small components $\kappa_i$ and $\hat{\kappa}_i$.
 The transverse momentum
 integrals in eq.~(\ref{factorizedA}),
 however, are linked by delta functions.
The transverse delta functions in the factorized form of the amplitude
 reflect the physical picture of p-p scattering,
 as it procedes in the multiple scattering mechanism.
   The incoming and the outgoing proton wave functions
 are highly Lorentz contracted along their respective light cone directions.
 Therefore, thought of as discs, they intesect along the direction
 perpendicular to the scattering plane \cite{BS}.  The hard scatterings may
 take place anywhere along this line of intersection.  Thus,
 in configuration space the wave
 functions are unlocalized in the normal to the scattering plane.
 When their separations are large, soft gluons moving in the
 normal direction can resolve the color of the quarks.  These
 effects are summarized in $U$.
   The amplitude is conveniently represented by rewriting
 eq.~(\ref{factorizedA}) in ``impact parameter" space in the
 transverse components,
 \begin{eqnarray}
{\cal A}(s,t,h_{i}) &=& (-i)\frac{(2\pi)^{6}}{\sin^{2}\theta}
   \int_{0}^{1} \d\xi_1 \d\xi_2
   \int \d b_1\d b_2    \nonumber \\
&\ & \times \sum_{f=1}^{3} \sum_{ \stackrel{\alpha_{i},}{a_{i},}
    \stackrel{ \beta_{i},} {b_{i},}\stackrel{\gamma_{i}}{c_{i}} }
    {\tilde Y}^{(f)}_{ \alpha_{i} \beta_{i}\gamma_{i}}
    (\xi_1,\xi_2, b_1,b_2;P_{i},h_{i}) \nonumber \\
&\ & \times {\tilde U}_{\{ a_{i} b_{i} c_{i}\} }(b_1,b_2,\mu)
\prod_{m=1}^3 H_{\{\alpha_i,a_i\}}(\xi_i,P_i,\mu)\, ,
\label{factorizedbspace}
\end{eqnarray}
 where $\mu$ is the renormalization scale
 and where $\tilde{Y}$, defined by
 \begin{eqnarray}
 \tilde{Y}^{(f)}_{ \alpha_{i} \beta_{i}\gamma_{i}}
    (\xi_1,\xi_2, b_1,b_2;P_{i},h_{i})
    &=&
    \int \d \kappa_i\d \hat{\kappa}_i \int \d^2 {\bf k}_i \d^2 {\hat {\bf k}}_i
    e^{-i({\bf b}_1\cdot {\bf k}_1+{\bf b}_2\cdot{\bf k}_2)} \nonumber \\
    &\ & \times Y^{(f)}_{ \alpha_{i} \beta_{i}\gamma_{i}}
    (k_{i},\hat{k}_{i};
    P_{i},h_{i})\, ,
    \label{tildeYdef}
    \end{eqnarray}
 is $Y$ integrated over the
 $\kappa$'s and Fourier transformed with respect
 to its transverse momentum components, in terms of the vectors
 ${\bf b}_i=\b_i\eta$, with $\eta$ the normal to the scattering plane.
 This is the factorized expression we shall employ below.

 Let us now discuss the various functions in eq.\ (\ref{factorizedA}) in
 more detail.

 \subsection{Wave functions on and off the light cone}

 Before assigning quarks to the independent scatterings we review the
 description of the ($uud$) Fock state.
 We omit the flavor flow index $(f)$ in $Y$ until we reinsert it in
  ${\cal A}$, eq.\ (\ref{factorizedA}).
  It is convenient to discuss the wave functions first in momentum space.

 As usual, the three-quark wave function is the Fourier
 transformation of the three-quark operator \cite{CZ}.
 For a proton moving in the + direction, with energy $E$,
 we have
\beqa
Y_{\alpha \beta \gamma}(k_{u_1},k_{u_2};P,h,\mu)
 &=& \frac{(2 E^2)^{1/4}}{N_{c}!}
\int
\frac{\d^{4}y_{1}}{(2\pi)^{4}} \e^{i k_{u_1} \cdot y_{1}}
\frac{\d^{4}y_{2}}{(2\pi)^{4}} \e^{i k_{u_2} \cdot y_{2}} \nonumber \\
&\ & \times \langle 0 | T [u^{a}_{\alpha}(y_{1}) u^{b}_{\beta}(y_{2})
              d^{c}_{\gamma}(0) ] |P,h \rangle
\epsilon_{a b c}\, .
\label{tensorY}
\eeqa
In the limit $y_i^\mu\rightarrow y_i^+v^\mu$, the three operators approach the
light cone, and $Y$ is related to a normal light-cone wave function \cite{CZ}.
$\mu$ is the factorization scale.  As mentioned above, the $Y$'s are
defined as color-invariant functions (appropriate ordered exponentials
between the quark fields are suppressed).

 We have already seen that
  the fields in eq.\ (\ref{tensorY}) are not on the light cone
 in general.  Nevertheless, the ratio of transverse to longitudinal
 momenta of the quarks will, in all the wave functions discussed
 below, vanish in the high-energy limit.  As a result,
 the leading twist contribution to the amplitude,
 eq.\ (\ref{factorizedA}),
 may  still be given by a
 spinor basis on the light cone.

 Permutation symmetry between the two $u$-quarks and the requirement of
 total isospin 1/2 imply that
 $Y_{\alpha\beta\gamma}$ can
 be expanded directly in terms of products of spinors
 with definite helicity,
\beqa
&\ &{\cal M}^{(1)}_{\alpha \beta \gamma} =
 (E_1 E_2 E_3 /2)^{-1/2} \;
 u_\alpha(k_{u_1},+) \; u_\beta(k_{u_2},-)  \; d_\gamma(k_d,+) \, ,
\nonumber \\
&\ &{\cal M}^{(2)}_{\alpha \beta \gamma} =
 (E_1 E_2 E_3 /2)^{-1/2} \;
 u_\alpha(k_{u_1},-) \; u_\beta(k_{u_2},+) \; d_\gamma(k_d,+) \, ,
\nonumber \\
&\ &{\cal M}^{(3)}_{\alpha \beta \gamma} =
- (E_1 E_2 E_3 /2)^{-1/2} \;
 u_\alpha(k_{u_1},+) \; u_\beta(k_{u_2},+) \; d_\gamma(k_d,-)  \, ,
\label{helspinors}
\eeqa
  as
\beq
Y_{\alpha \beta \gamma} = \frac{1}{2^{1/4} 8 N_c!}  f_N(\mu)
\left[
\Psi_{123} {\cal M}^{(1)}_{\alpha \beta \gamma}
+ \Psi_{213}   {\cal M}^{(2)}_{\alpha \beta \gamma}
+ \left( \Psi_{132}+\Psi_{231}\right )  {\cal M}^{(2)}_{\alpha \beta \gamma}
\right] \, ,
\label{Yhel}
\eeq
where the subscripts on the scalar
function $\Psi$ refer to the order of
momentum arguments, with $k_3=k_d$, for
example, $\Psi(k_1,k_2,k_3)\equiv \Psi_{123}$. In (\ref{helspinors}),
$E_1,\, E_2$ and $E_3$ are the energies of the first and second
 $u$-quarks and the $d$-quark respectively. For a proton with $(-)$
 helicity, it is only
 necessary to flip the spinor helicities
 in eq.\ (\ref{helspinors}). Here and below, spinors are
 normalized according to
 \beq
\overline{N}(P,h) \gamma^\mu N(P,h) = 2 P^\mu \, .
\label{spinornorm}
\eeq
A commonly-used alternate basis for $Y$ is described in the Appendix.

 We are now ready to discuss the
  transverse structure of the three quark state, since
  this is crucial to the factorized amplitude as given in
  eq.~(\ref{factorizedbspace}).
  At high energy, the dominant radiative corrections come
  from soft gluons, whose couplings are {\it indpendent
  of the quarks' spin states} \cite{BS}.  The effects
  of these gluons on the wave functions factorize
  from the light cone quark distribution amplitude
  (referred to as LCDA below)
  and exponentiate in the impact
  parameter ($b$) space respresentation.
That is, ${\tilde Y}$, eq.~(\ref{tildeYdef}),
the Fourier transform of the
 the function $\Psi(k_{u_1},k_{u_2},k_{d};\mu)$
may be written as
\beqa
{\tilde Y}_{\alpha\beta\gamma}(\xi_1,\xi_2,b_1,b_2;P_i,h_i)
&=&
\frac{1}{2^{1/4} 8 N_c!}  f_N(\mu)
\bigg [
{\cal P}_{123} {\cal M}^{(1)}_{\alpha \beta \gamma} \nonumber \\
&\ & \quad + {\cal P}_{213}   {\cal M}^{(2)}_{\alpha \beta \gamma}
+ \left( {\cal P}_{132}+
{\cal P}_{231}\right )  {\cal M}^{(2)}_{\alpha \beta \gamma}
\bigg ] \, ,
\label{tildeYexpand}
\eeqa
where
 \beqa
{\cal P}(\xi_1,\xi_2,\xi_3,\b_1,\b_2;\mu)
&=& \phi(\xi_1,\xi_2,\xi_3;\mu)
\tilde{\psi}(\xi_1,\xi_2,\tilde{b}_1,\tilde{b}_2)
\nonumber \\
&=&
\int \d \kappa_1 \int \d \kappa_2
\int \d^2 \k_1 \e^{ -i \k_1 \cdot \eta (b_1-b_3) }
\int \d^2 \k_2 \e^{ -i \k_2 \cdot \eta (b_2-b_3) }
\nonumber \\
&\ & \quad \times
\Psi(k_1,k_2,k_3;\mu)\, ,
\label{imprep}
\eeqa
with $\tilde{\psi}$ is a dimensionless function of the transverse
separations.  To be specific,
 we denote by $b_m$ the positions of the
 hard scatterings $H^m$ along the $\eta$ direction
 and by $\b_m$ their  mutual transverse separations defined as
\beq
\b_1=b_2-b_3 \, ,
\hspace{1cm}
\b_2=b_1-b_3 \, ,
\hspace{1cm}
\b_3=\b_2-\b_1 \, .
\label{btilde}
\eeq
 Another connection between the function ${\cal P}$ and the LCDA $\phi$ is
 \cite{BS}
\beq
{\cal P}(\xi_1,\xi_2,\xi_3, \b_1=\b_2=0; \mu) =
\phi(\xi_1,\xi_2,\xi_3;\mu)+{\cal O}\left (\alpha_s(\mu)\right) \, ,
\label{Pphi}
\eeq
that is, neglecting perturbative corrections for large $\mu$,
\beq
{\tilde \psi}(\xi_1,\xi_2,0,0)=1\, .
\label{tildepsinorm}
\eeq
Alternately, in momentum space, $\Psi$
is related to the LCDA by
\beq
\int \d \kappa_1 \int \d \kappa_2
\Psi(k_{u_1},k_{u_2},k_d;\mu) =
\phi(\xi_1,\xi_2,\xi_3;\mu) \psi(\xi_1,\xi_2,\xi_3,\k_{u1},\k_{u2}) \, ,
\label{Fourierpsi}
\eeq
with
\beq
\int \d^2{\bf k}_1 \d^2{\bf k}_2\; \psi(\xi_1,\xi_2,\xi_3,\k_1,\k_2;\mu)
=1\, .
\eeq

At very high energy, the behavior of ${\cal P}$ in the ${\tilde b}_i$
is a computable exponential \cite{BS}, which {\it vanishes} when
any of the ${\tilde b}_i$ equal $1/\Lambda$.   The perturbative
transverse wave function, then, is of finite size.  Rather than
give its explicit form here, we have simply
absorbed it into the resummed soft-gluon factor specified in
eq.~(\ref{s}) below.  In the following, we shall assume that
this has been done, and consider the function $\psi(\xi_i,{\tilde b}_i)$
as a measure of the remaining ``nonperturbative" transverse
structure of the proton.  As we shall see in Sect.~5 below,
this structure has strong influence on near-forward elastic
scattering.  In a sense, we may think of $\psi$ as the source
of the ``intrinsic" transverse momentum of valence quarks in
the proton.  We should keep in mind, however, that the
distinction between perturbative and nonperturbative, or
process-dependent and intrinsic, is somewhat arbitrary,
and must depend, in particular, on the order to which the perturbation
expansion has been carried out.

As a last topic in this subsection, we briefly review the evolution of
the light cone distributions.
 In the standard analysis \cite{BL},
 the LCDA is expanded in the basis of the Appel polynomials $A_j$  as
\beq
\phi(x_1,x_2,x_3;\mu) = \phi_{\rm as}(x_1,x_2,x_3) \sum_{j=0} N_j
\left( \frac{\ln \mu^2/ \Lambda^2}{\ln \mu^2_0/ \Lambda^2} \right)
^{-\frac{b_j}{ \beta_1}} a_j A_j(x_1,x_2,x_3) \, .
\label{Appel}
\eeq
 $\phi_{\rm as}(x_1,x_2,x_3)=120 x_1x_2x_3$ is the asymptotic limit
 of $\phi(\mu)$ for $\mu \rightarrow \infty$. $\mu_0 \approx$ 1GeV,
 $\beta_1=(11/3)N_c-(2/3)n_f$ is the one-loop coefficient of the QCD
 $\beta$-function and $\Lambda \equiv \Lambda_{\rm QCD}$
 is the QCD scale parameter.
 The normalization coefficients $N_j$, defined by
\beq
N_j\int_0^1\d x_1\d x_2 \d x_3
\delta(1-x_1-x_2-x_3)\;  \phi_{as}(x)A_j^2(x)=1 \, ,
\label{Nj}
\eeq
 and the anomalous dimensions $b_j/\beta_1$ are the same for all models.
 Model dependence of $\phi$ comes through the coefficients $a_j$.
 For a list of the constants and functions
 $N_j \, , b_j \, , a_j \, , A_j(x)$ see table 2 and
 \cite{Hyer,JSLN}.
 The dimensional parameter $f_N(\mu)$ is given by
\beq
f_N(\mu) = f_N(\mu_0)
\left( \frac{\ln \mu^2/ \Lambda^2}{\ln \mu^2_0/ \Lambda^2} \right)
^{-\frac{2}{3 \beta_1}}\, ,
\hspace{1cm} f_N(\mu_0)=(5.2 \pm 0.3)\times 10^{-3}{\rm GeV}^2 \, .
\label{fN}
\eeq
 Both $\phi$ and $f_N$ depend mildly on $\mu$ due to QCD evolution.

 The quark distribution amplitudes identified above
 contain much of the soft physics of the problem. However, information
 about soft gluon exchange is also contained in the
 perturbative resummation
 factors. As mentioned above, these will include the
 perturbative $b$ dependence of the wave functions \cite{BS}.
 To organize this dependence, we must analyze color flow in hard and
 soft exchange.

 \subsection{Color flow, hard and soft exchange}

 According to the conventions of fig.\ 1,
 $k_i,\hat{k}_i,P_i-k_i-\hat{k}_i$
 are the momenta entering ($i=1,2$) or leaving ($i=3,4$) the blobs
 $H^1,H^2$ and $H^3$ respectively.
 The identification of these momenta with
 $(k_{u_1})_i, (k_{u_2})_i, (k_d)_i$,
 is obtained once the flavor flow $f$ is fixed.
 Similarly, the third index $\gamma$ of the proton wave function,
 eq.\  (\ref{tensorY}), is identified  with the
 Dirac index $\beta_{i}$ or $\gamma_{i}$ in eq.\ (\ref{factorizedA})
 according to the hard scattering $H^m$, $m=2,3$, in which
 the $d$-quark of  the corresponding proton participates.
Each hard-scattering function $H^m$ consists of lines with virtuality
 $\xi_m^2 t$, with
 $\xi_m, \, m=1,2,3$, the longitudinal momentum fraction
 carried by the
 quarks in scattering $m$ (they are all the same for a given $m$).

 The color flow basis suitable for our
 treatment is the octet-singlet basis,
 \beqa
 H^m_{\{\alpha_i,a_i\}}
 &=&
 \left (H^m_{\rm s}\right )_{\{\alpha_i\}}\left ( c_{\rm s} \right )_{a_i}
 +
 \left (H^m_{\rm s}\right )_{\{\alpha_i\}}\left ( c_{\rm s} \right )_{a_i}\, ,
 \nonumber \\
 \left( c_{ {\rm adj}} \right)_{\{ a_i \}} &\equiv&
  (T_m)_{a_3 a_1} (T_m)_{a_4 a_2}\, ,
\hspace{1cm}
\left( c_{ {\rm s}} \right)_{\{ a_i \}} \equiv \delta_{a_1 a_3} \delta_{a_2
a_4}
\, .
\label{basis}
\eeqa
 For near-forward scattering the anomalous dimension matrix
 associated with $H^m$ becomes diagonal in this basis
 \cite{Lipquasel,SS1,Korch}. More detail will be given in
 the following section.

 The color tensor $U$ includes soft radiative corrections that
 cannot be absorbed into the
  proton wave functions. As such, its perturbative
  expansion begins at zeroth order in the coupling,
\beq
{\tilde U}_{ \{a_{i} b_{i} c_{i}\} }(b_1,b_2,\mu) = \prod_{i=1}^{4}
\epsilon_{ a_{i} b_{i} c_{i} }+ {\cal O}(\alpha_{s}) \, .
\label{U}
\eeq
At lowest order, $U$ consists simply of the $\epsilon$
color tensors associated with the external protons.
 In ref. \cite{BS} the one-loop radiative corrections for fixed angle p-p
 scattering were analyzed, and the
 corresponding renormalization group equations identified.
  The IR safety of the p-p process was shown to arise from the
 cancellation of IR divergences among graphs of the form shown in fig.\ 2.
 The underlying physical principle is that there is an effective IR cut-off
 $1/R$  for the transverse momentum of the exchanged gluons,
 with $R$ the distance between adjacent $H^m$'s.
 Below this cut-off, long wavelength gluons cannot couple to
 the color singlet hadronic state.

 Leading logarithms in $s/\mu^2$, $t/\mu^2$ and $b_i^2\mu^2$ may be
 resummed into exponential facotrs, $\exp(-s_I)$,
  one for each hard scattering function $H^m_I$.  As above,
  $I=\rm s,\; adj$ labels the color flow of $H^m$ in the basis
  (\ref{basis}).  Beyond one loop in the corresponding
  anomalous dimensions, the exponentials for each
  hard scattering no longer factorize from each other in general.
  Once leading logarithms have been organized in this
  fashion, the renormalization scale $\mu$ in
  $H^m$ is replaced by the hard-scattering scale
  \beq
  Q_m=\xi_m C_2\sqrt{-t}\, ,
  \label{Qmdef}
  \eeq
  where $C_2$ is a parameter chosen to optimize
  higher order corrections.  Similarly, in ${\tilde U}$, $\mu$ may be
  replaced by a scale of the order of the $1/{\tilde b}_i$'s.  In
  the following, we shall neglect higher orders
  in $\alpha_s(1/{\tilde b}_i)$, and retain only the first term in
  the expansion of ${\tilde U}$, eq.~(\ref{U}).

 The exponents that are generated by the one-loop radiative corrections
 to the hard scattering $H^m$,  with color flow $I$, are given by \cite{BS}
\beqa
 s_{_I}(Q_m, \b_m) &=& \quad
c_1 \left[ q_m \ln \left( \frac{q_m}{\zeta_m} \right) -q_m +\zeta_m \right]
+ c_2 \left[ \frac{q_m}{\zeta_m} -\ln \left( \frac{q_m}{\zeta_m}\right)
  -1 \right]  \nonumber \\
&\ & + \, c_3 \left[ 2 ( \ln(2q_m)+1)-2 \frac{q_m}{\zeta_m}(\ln(2\zeta_m)+1)
                 -\ln^2(2 q_m) + \ln^2(2 \zeta_m) \right] \nonumber \\
&\ & + \, \frac{2}{\beta_1} B_{_I}\ln \left( \frac{q_m}{\zeta_m}\right) \, ,
\label{s}
\eeqa
with
\beq
q_m \equiv \ln(Q_m/\Lambda)=\ln(\xi_m C_2 \sqrt{-t}/\Lambda) \, ,
\hspace{1cm}
\zeta_m\equiv-\ln(|\b_m|\Lambda) \, .
\label{qzeta}
\eeq
 The positive constants $c_1$, $c_2$, $c_3$, $\beta_1$ and $\beta_2$ are
\beqa
&\ &c_1=\frac{8 C_F}{\beta_1} \, ,
\hspace{1cm}
c_2= \frac{8}{\beta_1^2} \left[
N_c C_F \left( \frac{67}{18} -\zeta(2) \right) -\frac{5}{9} C_F n_f
 + C_F \beta_1 (\gamma_E-\ln2) \right] \, ,
\nonumber \\
&\ & c_3 = \frac{C_F \beta_2}{2 \beta_1^3} \, ,
\hspace{0.8cm}
\beta_1= \frac{11}{3}N_c - \frac{2}{3}n_f \, ,
\\
&\ &\beta_2 = \frac{34}{3} N_c^2
- \left( \frac{20}{3} N_c^2 + 4 C_F \right) T_F n_f \, .
\nonumber
\label{constants}
\eeqa
 For $N_c=3$ and normalization $T_F=1/2$, we have $C_F=4/3$ and we
 take $n_f=3$.
 $\beta_2$  is  the two-loop coefficient of the QCD $\beta$-function
 and $\gamma_E$ is the Euler constant.
 Of particular interest is the last term $B_{_I}$
 in eq.\ (\ref{s}), which contains the color flow information of the hard
 scattering,
\beq
 B_{_I}=\lambda_{_I}+2C_F \ln \left( \frac{\e^{2 \gamma_E-1}}{4 C_2^2} \right)
 -N_c C_F \, .
\label{Beta}
\eeq
 Here, $\lambda_{_I}$ is the eigenvalue of the anomalous dimension matrix of
 $H^m$ along the color flow direction $I$, and the term $-N_c C_F$ comes from
 the anomalous dimensions of the participating quarks,
 $4 \gamma_q=-N_c C_F \alpha_s/\pi$ in the  axial gauge.
 The $\lambda_I$ will be specified in the next section.

To summarize our results so far, the p-p elastic amplitude may be
expressed in terms of integrals in $b$-space in the general form \cite{BS}
\begin{eqnarray}
{\cal A}(s,t,h_{i}) &=& (-i)\frac{(2\pi)^{6}}{\sin^{2}\theta}
   \int_{0}^{1} \d\xi_1 \d\xi_2
   \int \d {\tilde b}_1\d {\tilde b}_2    \nonumber \\
&\ & \times \sum_{f=1}^{3} \sum_{ \stackrel{\alpha_{i},}{a_{i},}
    \stackrel{ \beta_{i},} {b_{i},}\stackrel{\gamma_{i}}{c_{i}} }
    {\tilde Y}^{(f)}_{ \alpha_{i} \beta_{i}\gamma_{i}}
    (\xi_1,\xi_2, b_1,b_2;P_{i},h_{i}) \nonumber \\
&\ & \times U_{\{ a_{i} b_{i} c_{i}\} }({\tilde b}_i,\,\alpha_s(1/{\tilde
b}_i))
\left ( c_{I_1} \right)_{a_i}
\left ( c_{I_2} \right)_{b_i}
\left ( c_{I_3} \right)_{c_i} \nonumber \\
&\ & \times \exp\left ( -s_{I_1}(Q_1,{\tilde b}_1)
-s_{I_2}(Q_1,{\tilde b}_2)-s_{I_3}(Q_1,{\tilde b}_3) \right )
\nonumber \\
&\ & \times \prod_{m=1}^3
\left ( H_{I_m}\right )_{\{\alpha_i\}}(\xi_i,P_i,\alpha_s(Q_m))\, .
\label{rgimprovedamp}
\end{eqnarray}

\section{The quark-quark hard amplitude in the near-forward region}

 In the kinematic region of near forward scattering with large momentum
 transfer,
\beq
s \gg -t \gg \Lambda^{2} \, ,
\label{region}
\eeq
 the leading power contribution to the p-p elastic amplitude
 comes from the ``$ttt$''gluon exchange channel for $f=1$ and $f=2$
 (see table 1).
 The information about color flow in each hard scattering $H^m$ enters
 through the eigenvalue $\lambda_{_I}$ in the $B_{_I}$ term of the
 exponent, eq.\ (\ref{s}). In \cite{SS1} we have calculated the one-loop
 octet and singlet eigenvalues. The result is
\beq
\lambda_1 \equiv \lambda_{{\rm adj}} = N_c \ln \left( \frac{s}{-t} \right)
-\frac{2 \pi i}{N_c} + \lambda_\s \, ,
\hspace{1cm}
\lambda_\s = 2 C_F,
\label{eigenvalues}
\eeq
giving
\beq
B_{{\rm adj}}-B_\s= N_c \ln \left( \frac{s}{-t} \right)
+{\cal O}(\ln^0s) \, .
\label{Bdiff}
\eeq
 Consequently, the contribution to the p-p amplitude from octet exchange
 in a single hard scattering is suppressed relative to singlet exchange
 by a factor
\beq
\left( \frac{s}{-t} \right)^{-\frac{ 2N_c}{\beta_1} \ln(q_m/\zeta_m)} \, ,
\label{regge}
\eeq
 with $q_m$, $\zeta_m$ defined in eq.\ (\ref{qzeta}).
 We note that the $\lambda_I$ in eq.~(\ref{eigenvalues})
 are gauge dependent, but that the full result in eq.~(\ref{s})
 is gauge invariant \cite{BS}.
 Due to eq.\ (\ref{regge}),
 we expect that the dominant contribution to ${\cal A}$ in the
 near-forward region comes from singlet exchange in all
 three hard scatterings, which we denote by ${\cal A}_{3{\rm s}}$.

 For the octet, $H^m_{\rm adj}$
 begins at
 single gluon exchange, but the lowest order contribution
 to the singlet, $H^m_{\rm s}$ is
 at the one-loop level,
  and the contributing on-shell graphs are IR divergent.
 The IR finite hard part of
 $H^m_{\rm s}$ can be put in the form \cite{SS1}
\beqa
H^{m(1)}_{\s \{ \alpha_{i} \}} \left(c_\s \right)_{\{a_{i}\}} &=&
\tau^m(\alpha_s(Q_m^2),\xi_m^2t)
(\gamma^\mu)_{\alpha_3 \alpha_1}
(\gamma_\mu)_{\alpha_4 \alpha_2} C_\s \left(c_\s \right)_{\{a_{i}\}} \, ,
\nonumber \\
\tau^m(\alpha_s(Q_m^2),\xi_m^2t) &=&
\frac{4 \pi \alpha_{s}^2(Q_m^2)}{\xi ^{2} t}
 \ln \left( \frac{Q_m^2}{\xi_m^2 \overline{Q}^{2}} \right) \, ,
\label{lowhard}
\eeqa
where
\beq
C_\s=\frac{N_c^2-1}{4N_c^2}
\label{colorcoeff}
\eeq
 is the color coefficient coming from the projection of the
 lowest order color structures along the $t$-channel color singlet, and
 where we define
\beq
\overline{Q}^{2}= \frac{\e^{\gamma_{_{E}}-\rho}(-t)}{4 \pi} \, .
\label{rhointro}
\eeq
 $\rho$ parameterizes the ambiguity in the
 IR subtraction to lowest order.
 We expect an optimal $\rho$ to be determined by an examination of higher
 orders.

 Inspection of eqs.\ (\ref{Qmdef}), for $Q_m$, and (\ref{lowhard}),
 for $\tau_m$, shows that an ``$\overline{\rm MS}$"
 choice for the parameter $C_2$,
\beq
C_2^2=\frac{\e^{\gamma_E}}{4 \pi} \, ,
\label{C2MSbar}
\eeq
 gives a hard singlet that is proportional to $\rho$.
 We take for $C_2$ this $\overline{\rm MS}$ choice
 for the rest of this paper and denote
\beq
\overline{\tau}^m(\alpha_s(\overline{Q}_m^2), \xi_m^2t)
 =-i \frac{4 \pi \alpha_s^2(\overline{Q}_m^2)}{\xi^2_m t} \rho \, .
\label{taubar}
\eeq

 With the choice (\ref{C2MSbar}) for $C_2$, the $B_\s$ coefficient
 of the Sudakov exponent,  eq.\ (\ref{Beta}),
 using eq.\ (\ref{eigenvalues}) for $\lambda_\s$, becomes
\beq
B_\s=2 C_F (\gamma_E + \ln \pi) - N_c C_F \, .
\label{Betabar}
\eeq

\section{Calculation of the p-p amplitude}

 At this stage, all the necessary ingredients are available for the
 calculation of ${\cal A}_{3{\rm s}}$.
  First we calculate the lowest order $(\alpha_s^6)$ contribution
 to the hard-scattering amplitude.
  The calculation is straightforward using
   eqs.\ (\ref{helspinors}), (\ref{Yhel}),
 and (\ref{spinornorm}).
  To present the result in a compact form we define
\beq
{\cal T}_{123} \equiv \frac{1}{2} [ {\cal P}_{132} +{\cal P}_{231} ]\, ,
\label{psiphi}
\eeq
 for each external wave function, with ${\cal P}$ defined
 in eq.~(\ref{imprep})
 . The result in the
 impact parameter representation, eq.\ (\ref{rgimprovedamp}), is found from
\beqa
&\ &
\sum_{f=1}^{2} \sum_{ \stackrel{\alpha_{i},}{a_{i},}
    \stackrel{ \beta_{i},} {b_{i},}\stackrel{\gamma_{i}}{c_{i}} }
    {\tilde Y}^{(f)}_{ \alpha_{i} \beta_{i}\gamma_{i}}
    (\xi_1,\xi_2, b_1,b_2;P_{i},h_{i})
{\tilde U}_{\{ a_{i} b_{i} c_{i}\} }({\tilde b}_i,\,\alpha_s(1/{\tilde b}_i))
\left ( c_{I_1} \right)_{a_i}
\left ( c_{I_2} \right)_{b_i}
\left ( c_{I_3} \right)_{c_i} \nonumber \\
&\ & \quad\quad\quad\quad \times \prod_{m=1}^3
\left ( H_{I_m}\right )_{\{\alpha_i\}}(\xi_i,P_i,\alpha_s(Q_m))
\nonumber \\
&\ & = -
\overline{\tau}^1
\overline{\tau}^2
\overline{\tau}^3 {\cal C}\delta_{h_1 h_3} \delta_{h_2 h_4}\;
f_N^4(\mu)
\frac{1}{4}
\bigg\{ [ {\cal P}_{123}^2+ {\cal P}_{213}^2 + 4 {\cal T}_{123} ^2 ]^2
\nonumber \\
&\ & \quad\quad\quad\quad\quad \quad  +
        [ {\cal P}_{123}^2+ {\cal P}_{213}^2 + 4 {\cal T}_{123} ^2 ]
        [ {\cal P}_{132}^2+ {\cal P}_{312}^2 + 4 {\cal T}_{132} ^2 ] \bigg\}
\, .
\label{tree}
\eeqa
 Notice that the first term inside the braces, from $f=1$, is symmetric under
 $1 \leftrightarrow 2$, because for $f=1$,
 $H^1$ and  $H^2$ represent the same flow of flavor.

 The color factor ${\cal C}$ is determined by the normalization $N_c !$ of the
 $Y$'s, eq.~(\ref{tensorY}), the eikonal tensor ${\tilde U}$, eq.\ (\ref{U}),
 and the color structure of the $H^m$'s, eqs.\ (\ref{lowhard}),
 (\ref{colorcoeff}). It is
\beq
{\cal C} =  \frac{ C_\s^3}{(N_c!)^4} \left( \prod_{i=1}^4
 \epsilon_{ a_i b_i c_i} \right)
 \left( c_\s \right)_{ \{a_i \}} \left( c_\s \right)_{ \{b_i \}}
 \left( c_\s \right)_{ \{c_i \}} \nonumber \\
=\left( \frac{N_c^2-1}{4 N_c^2} \right) ^3 \frac{1}{(N_c!)^2} \, ,
\label{colorfactor}
\eeq
 yielding ${\cal C}=2/3^8$ for $N_c=3$.

 Including the soft-gluon factors for singlet exchange and  using \\
 $\sin^2\theta=-4t/s(1+{\cal O}(t/s))$ we obtain
\beqa
{\cal A}_{3{\rm s}}(s,t) &=& -\frac{1}{(4\pi)^2} \frac{s}{t}
\int_0^1 \d \xi_1 \d \xi_2
\int_{-\infty}^\infty \d \b_1 \d \b_2
[T_{3{\rm s}}]^{(1)}
{\cal R}(\xi_1,\xi_2, \b_1, \b_2;\mu)
\nonumber \\
&\ & \quad \times \exp[ -s_\s(\overline{Q}_1,|\b_1|)
                        -s_\s(\overline{Q}_2,|\b_2|)
                        -s_\s(\overline{Q}_3,|\b_1-\b_2|)] \, ,
\label{A3P}
\eeqa
 with the lowest order contribution $[T_{3{\rm s}}]^{(1)}$ given by
\beqa
[T_{3{\rm s}}]^{(1)} &=& i(2 \pi)^8
\overline{\tau}^1 \overline{\tau}^2 \overline{\tau}^3 {\cal C}
\delta_{h_1 h_3} \delta_{h_2 h_4}
\nonumber \\
&=& -\frac{2}{3^8} (2 \pi)^8
\frac{ (4 \pi)^3
       \alpha_s^2(w_1^2)
       \alpha_s^2(w_2^2)
       \alpha_s^2(w_3^2)}{\xi_1^2 \xi_2^2 (1-\xi_1 -\xi_2)^2 t^3} \rho^3
\delta_{h_1 h_3} \delta_{h_2 h_4}\, .
\label{T3P}
\eeqa
 The second line in the above equation follows from eq.\ (\ref{taubar}), which
 defines the hard scatterings,
 and we denote by $w_m$ the hard scale at which
 each hard scattering $H^m$ is evaluated.
 The wave function combination ${\cal R}$ is
\beqa
{\cal R} &=& f_N^4(\mu) \frac{1}{4}
\left\{
[ {\cal P}_{123}^2(\mu)+ {\cal P}_{213}^2(\mu) + 4 {\cal T}_{123} ^2 (\mu) ]^2
  \right. \nonumber \\
&\ & + \left. [ {\cal P}_{123}^2(\mu)+ {\cal P}_{213}^2(\mu)
+ 4 {\cal T}_{123} ^2 (\mu) ]
              [ {\cal P}_{132}^2(\mu)+ {\cal P}_{312}^2(\mu)
+ 4 {\cal T}_{132} ^2 (\mu) ]
\right\} \, .
\label{R}
\eeqa
 In summary, the
 three singlet amplitude is given by
 eqs.\ (\ref{A3P})-(\ref{R}), with soft
 gluon resummation exponents
 for singlet exchange given in eqs.\ (\ref{s}) and (\ref{Betabar}).
 These results are the same for proton-antiproton near-forward
 hard elastic scattering, since the corresponding amplitude
 is obtainable from $s \leftrightarrow u$  crossing of p-p scattering,
 and in the forward direction $u^2 \approx s^2$.

\section{Numerical results}

 The experimental study \cite{ISR,FNAL} of proton-proton near-forward elastic
 scattering reveals two characteristic features
 in the momentum transfer range of interest here. The first is the
 independence of the differential cross section $\d \sigma^{pp}/ \d t$
 from the energy, for an energy range $\sqrt{s} \ge 25$ GeV up to
 62 GeV.
 The second is the power behavior $\d \sigma^{pp}/ \d t \sim t^{-8}$.
 The lowest order multiple scattering model \cite{Land} predicts precisely
 this scaling and power behavior.
 Landshoff and Donnachie \cite {DonnLand}
 were able to fit the experimental data in the range
 $ 3 \, {\rm GeV}^2 \le -t \le 15 \,  {\rm GeV}^2$ by
\beq
\frac{\d \sigma^{pp}}{\d t} = C t^{-8} \, ,
\hspace{1cm}
C=0.09 \, {\rm mb (GeV)}^6 \, .
\label{Landfit}
\eeq
 First, we examine the
 factorized amplitude eq.~(\ref{rgimprovedamp})
 in this region
  without
 transverse structure for the wave functions beyond the
 soft gluon exponentials $\exp[-s_{\rm s}]$,
 that is, with $\psi(\xi_i,{\tilde b}_i)=1$ in eq.~(\ref{imprep}).
 Then we shall include a model
 for nonperturbative transverse structure and study
 its influence.

 \subsection{Perturbative transverse structure}

 Upon retaining only the LCDA's in the proton wave-function
 (see eq.\ (\ref{Pphi})), the function ${\cal R}$ eq.\ (\ref{R}) becomes
\beqa
{\cal R} &=& f_N^4(\mu) \frac{1}{4}
\left\{
[ \phi_{123}^2(\mu)+ \phi_{213}^2(\mu) + 4 T_{123} ^2 (\mu) ]^2
  \right. \nonumber \\
&\ & + \left. [ \phi_{123}^2(\mu)+ \phi_{213}^2(\mu) + 4 T_{123} ^2 (\mu) ]
              [ \phi_{132}^2(\mu)+ \phi_{312}^2(\mu) + 4 T_{132} ^2 (\mu) ]
\right\} \, ,
\label{RnoXverse}
\eeqa
with $T_{123}\equiv (1/2)(\phi_{321}+\phi_{132})$.

 Energy-independence of the cross section follows from
 eqs.\ (\ref{A3P}), (\ref{T3P}) and the relation
\beq
\frac{\d \sigma}{\d t} = \frac{1}{16 \pi s^2} |\tilde{ {\cal A}}|^2 \, ,
\label{diffxsection}
\eeq
 in which the $1/s^2$ cancels with the $s^2$ in $|\tilde{{\cal A}}|^2$
 from eq.\ (\ref{A3P}).
 Also, the singlet eigenvalue $\lambda_\s$, appearing
 in the exponents $s_\s$, eqs.\ (\ref{Beta}), (\ref{eigenvalues}),
 is energy independent, unlike the octet,
 $\lambda_{ {\rm adj}}$. Exact $s$-independence is not anticipated
 to persist at very high energies, where the energy dependence of the
 singlet exchange hard scatterings should become important \cite{BFKL}.

 The $t^{-8}$-dependence is characteristic of the multiple scattering model
 {\em if} we assume that the integral over the suppression factors
 $\exp(-s_{\rm s})$
 yields negligible $t$-dependence. Indeed, then ${\cal A}_{3{\rm s}} \sim
s/t^4$
 and, from eq.\ (\ref{diffxsection}),
 $\d \sigma^{pp}/ \d t \sim t^{-8}$.
 This power dependence cannot persist at high enough $t$, where
 the strong Sudakov suppression
 in the exponents leads to the onset of dimensional
 counting behavior of the form ${\cal A}_{3{\rm s}} \sim s/t^5$,
 and hence to
 $\d \sigma^{pp}/ \d t \sim t^{-10}$.

 In the numerical study that follows we plot the quantity
 $N^{-1}|{\cal A}_{3{\rm s}}| \, t^4/s$ as a function of $-t$.
 The dimensionless normalization coefficient $N$ is
\beq
N = (2 \pi)^8 (4 \pi)^7 \frac{1}{\beta_1^6} {\cal C}
 = 6.895 \times 10^4 \, .
\label{N}
\eeq
It has been defined
 without the factor $\rho^3$, and we use the one-loop
 running coupling  $\alpha_s(w_m^2) = 4 \pi /(\beta_1 \ln(w_m^2/\Lambda^2))$
 in $[T_{3{\rm s}}]^{(1)}$, eq.\ (\ref{T3P}).
 The numerical results for the amplitude ${\cal A}_{3{\rm s}}$ are
 in units of $|\rho^3|$.
 The straight line labeled `DL' in the figures
 corresponds to the fit of eq.\ (\ref{Landfit}).
 In order to compute ${\cal A}_{3{\rm s}}$ from eq.\ (\ref{A3P})
 we need to fix the $\b_1$, $\b_2$ integration limits and the scales
 $\mu, w_1, w_2, w_3$ . This we do as follows.

 {\em i}) Lower limits for $\b_1,\, \b_2$.
 The exponents $s_\s$ have a root at
 $\overline{Q}_m = 1/|\b_m|$ (see eq.\ (\ref{s})). We refer to the
 region $ 0 \leq |\b_m| \leq  1/ \overline{Q}_m$ as the small-$b$ region.
 Formally in this region the factor $\exp[-s_\s]$ in eq.\ (\ref{A3P})
 enhances  instead of suppressing. We take its value in the small-$b$ region
 to be 1, i.e.,
\beq
s_\s(\overline{Q}_m,|\b_m|) \rightarrow s_\s(\overline{Q}_m, |\b_m|)
\theta(\overline{Q}_m -1/|\b_m|)
\label{scut}
\eeq
 and integrate down to $|\b_m| \rightarrow 0$ \cite{LiS}.

 {\em ii}) Upper limits for $\b_1, \, \b_2$.
 $|\b_m|$ are strictly limited to be less  than $1/ \Lambda$,
 as past this value the Sudakov exponents diverge.
 We impose upper cut-off $b_c$
 in the $\b_1, \, \b_2$ integrals and examine the behavior
 of ${\cal A}_{3{\rm s}}$  as $b_c \rightarrow 1/\Lambda$.

 {\em iii}) For the scales $\mu \, , w_m$ we make the following choices.
 The factorization scale $\mu$,
 separating wave functions from the remainder of the amplitude,
 (eqs.\ (\ref{Appel}), (\ref{fN})),
 is taken equal to the minimum average
 transverse momentum of the valence quarks, corresponding to the maximum
 transverse separation between the hard scatterings,
\beq
\mu = {\rm min} \left\{ \frac{1}{|\b_1|}, \frac{1}{|\b_2|},
\frac{1}{|\b_1-\b_2|} \right\} \, .
\label{mufix}
\eeq
 In the soft quark region, $\xi_m \rightarrow 0$, the hard scales $w_m$
 shift from $\overline{Q}_m$ to $\mu$  as
\beq
w_m = {\rm max } \{\overline{Q}_m, \mu \} \, .
\label{wfix}
\eeq
 In addition to the conditions (\ref{mufix}), (\ref{wfix}),
 we impose a lower bound $\mu_{\rm min}$ for all the scales
\beq
 \mu\,, w_1\,, w_2\,, w_3 \geq \mu_{\rm min} \, ,
\label{lowerbound}
\eeq
 and examine the sensitivity of the amplitude as
 $\mu_{\rm min} \rightarrow \Lambda$.

 The result depends, in principle, on the values of $\mu_{\rm min}$ and
 $\Lambda$, the model for $\phi(x_i;\mu)$
 and the value of the undetermined parameter $\rho$ in eq.\ (\ref{T3P}).
 In this numerical study we take $|\rho|=1$, but it should be noted
 that the results for the amplitude scale with $|\rho|^3$.
 We start with
 $\mu_{\rm min} = 1 \, {\rm GeV}$ and the LCDA model
 of Ref.~\cite{CZnuclphys} and look at
 the rate of convergence of ${\cal A}_{3{\rm s}}$ as $b_c \rightarrow
1/\Lambda$,
 as well as its sensitivity to the choice of the scale parameter $\Lambda$.
 The results are shown in figs.\ 3, 4. Note that for
 $\Lambda = 0.1 \, {\rm GeV}$, $\mu_{\rm min} = 10 \Lambda$ corresponds to
 $b=1/\mu_{\rm min}=0.1/\Lambda$ and for $\Lambda = 0.2 \, {\rm GeV}$,
 $\mu_{\rm min} = 5\Lambda$ corresponds to
 $b=1/\mu_{\rm min}=0.2/\Lambda$. As the figures
 show, up to these values for $b_c$ the Sudakov suppression has hardly started
 affecting the amplitude. Therefore, in the region of the impact parameter
 space where Sudakov effects become important the coupling constant is
 effectively  fixed at $\alpha_s(\mu_{\rm min}^2)$
 due to condition (\ref{lowerbound}).

 We tentatively divide the $t$-range into three regions
 corresponding to small (a), intermediate (b), and large (c),
 absolute slopes.
 We observe approximate linear dependence of
 $\ln (|{\cal A}_{3{\rm s}}| \: t^4/s) $
 on $\ln \:(t)$ in each of the above regions.
 Table 3 lists the slopes of the integral for
 selected values of $b_c$. The slopes on the logarithmic plot correspond to
 `$t^{{\rm slope}}$' deviation of the amplitude from the $t^{-4}$ behavior.
 The onset of dimensional counting in region (c) is apparent,
 where strong Sudakov suppression
 stem the growth of the amplitude. Note also the relatively small
 deviations from the $t^{-4}$ behavior in region (a).  less
 pronounced in region (b), where the transition to dimensional counting
 occurs.

 The dependence of the amplitude on $\Lambda$ is also of interest.
 We observe that, for this choice of $\mu_{\rm min}$ and LCDA model,
 ${\cal A}_{3{\rm s}}$ decreases with increasing $\Lambda$ for
 $b_c \le 0.3/\Lambda$ while it increases past this value of $b_c$.
 This is due to the competition between two counteracting effects. The
 first is the decrease of the impact parameter space volume with increasing
 $\Lambda$ due to the Jacobian factor $1/\Lambda^2$ in
\beq
\int_0^{b_c} \d \b_1 \d \b_2 =
\frac{1}{\Lambda^2}
\int_0^{b_c \Lambda} \d \b_1^\prime \d \b_2^\prime \, ,
\label{rescaling}
\eeq
 noting that the integrand is a function of $b_c \Lambda$ only.
 The second is the increase of the coupling constant with $\Lambda$. From
 $\Lambda = 0.1$ GeV to 0.2 GeV, $\alpha_s(\mu_{\rm min}^2)$ increases
 by a factor $\sim 1.4$.
 This gives a substantial effect because the lowest order contribution to
 the amplitude is ${\cal O}(\alpha_s^6)$.
 In the case of the proton form factor \cite{Li},
 where the lowest order contribution is ${\cal O}(\alpha_s^2)$,
 the shrinking of the impact parameter space is always stronger than the
 increase of $\alpha_s$ with increasing $\Lambda$, even when
 $\mu_{\rm min} \rightarrow \Lambda$.
 Proton-proton near-forward  scattering is
 much more sensitive to changes in $\alpha_s$, even though
 the Sudakov suppression here is stronger relative to the form factor,
 due to the presence of the positive eigenvalue $\lambda_\s = 2C_F$
 (eq.~(\ref {eigenvalues})) in the
 exponent $s_\s$.

 Decreasing the lower bound $\mu_{\rm min}$ down to 0.7 GeV, we recompute
 the amplitude for the same combination of parameters as before. The
 results are shown in figs. 5, 6.
 Again here, for $b_c \ge 0.3/\Lambda$ the coupling
 is essentially fixed at $\alpha_s(\mu_{\rm min}^2)$.
 The main feature is the marked increase of the amplitude
 (notice the scale shift on the  ${\cal A}_{3{\rm s}}$ axis).
 For $\Lambda = 0.1$ GeV,  ${\cal A}_{3{\rm s}}$ increases by a factor
$\sim$3.0
 in region (a) and by a factor $\sim$2.3 in region (c).
 For $\Lambda = 0.2$ GeV the increase is by a factor $\sim$5.5 in (a) and
 $\sim$3.4 in (c). This shows that the calculation is quite sensitive to the
 lower bound $\mu_{\rm min}$.
 Although Sudakov suppression always makes the integral finite,
 removing the condition
 (\ref{lowerbound})
 ( equivalent to letting $\mu_{\rm min} \rightarrow \Lambda$ )
 leads to an increase  by orders of magnitude.

 For fixed $t$, the relative logarithmic distance between the $b_c$ curves
 remains almost constant with varying $\mu_{\rm min}$ at fixed $\Lambda$.
 This is a safe diagnostic that the dependence of the amplitude
 on $\mu_{\rm min}$ is due to the coupling constant, since for the
 largest part of the impact parameter space
 $\alpha_s=\alpha_s(\mu_{\rm min}^2) \propto \ln^{-1}(\mu_{\rm min}/\Lambda)$.
 The drift of ${\cal A}_{3{\rm s}}$ to higher values with decreasing
 $\mu_{\rm min}$ does not affect the slope of the $b_c$ curves in region (c).
 Dimensional counting is bound to set in for high enough $t$. But the
 transition region from approximate $t^{-4}$
 to dimensional counting behavior shifts
 towards smaller $t$ with decreasing $\mu_{\rm min}$.
 This is not due to any change in
 Sudakov effects, but to the fact that the smaller $t$ region receives larger
 corrections from $\alpha_s(\mu_{\rm min}^2)$.

 The general characteristics of the ${\cal A}_{3{\rm s}}$ behavior described
above
 are the same for all LCDA models, and the following conclusion can be drawn.
 Sudakov suppression is not strong enough by itself to stabilize
 the pQCD calculation of the p-p near-forward amplitude
 against contributions from soft regions.
 This is not to be interpreted as a failure, but rather as an indication of
 the importance of the
 transverse structure of the three-quark Fock state modeling
 the proton. It is apparent that in the $t$-region where $t^{-4}$ behavior
 is observed, the Sudakov suppression sets in at unphysical scales,
 larger than the transverse size of the proton.
 $t^{-4}$ behavior for the amplitude is due to ``geometrical" scattering,
 to which the entire size of the proton can contribute,
 and the transition  to dimensional counting occurs only once
 the Sudakov suppression is strong enough to confine
 the independent scatterings to regions smaller
 than the average transverse size of the three quark state. In this way
 we can assign physical meaning to the scale $\mu_{\rm min}$, as being of the
 order of the average transverse momentum of the valence quarks inside the
 proton. Following this reasoning, we proceed by including in the
 calculation the transverse structure of the proton. \\

\subsection{Intrinsic transverse structure}

 Returning to the factorized form of the hadronic wave function,
 eq.\ (\ref{Fourierpsi}),
\beq
\Psi(k_{u_1}, k_{u_2}, k_{d} ; \mu) =
\phi(x_1, x_2, x_3 ;\mu) \psi(x_i,\k_i) \, ,
\label{includepsi}
\eeq
 we consider a transverse wave function of the form
 \cite{BL,Banff,Dziem,JacKroll,Hyer}
\beq
\psi(x_i, \k_i) =
N \exp \left[ -R_{3q}^2 \sum_{i=1}^3 \frac{\k_i^2}{x_i} \right] \, ,
\label{psi}
\eeq
 where $N=R_{3q}^4/(\pi^2 x_1 x_2 x_3)$ and $R_{3q}$ is a mean
 impact parameter.
 The impact parameter representation of $\psi(x_i,\k_i)$ is defined in
 eq.\ (\ref{imprep}). The result is
\beq
\psi(x_i,\b_i) =
\exp \left[ -\frac{1}{4 R_{3q}^2}
(x_1 x_2 \b_3^2 + x_1 x_3 \b_2^2 + x_2 x_3 \b_1^2)
\right] \, .
\label{impactpsi}
\eeq
 The value of $N$ is fixed by the requirement $\psi(\b_i=0)=1$
 (eq.~(\ref{tildepsinorm})).
 In this paper we do not advocate any particular choice of $\psi(x_i, \k_i)$,
 but use (\ref{psi}) because it has received much previous attention.

 The assignment of quarks to hard
 scatterings according to the flavor flow $f$
 does not affect $\psi(x_i,\b_i)$, because it is totally symmetric in its
 indices. The expression of the amplitude, eq.\ (\ref{A3P}), can be
 readily obtained in this case from ${\cal R}$, eq.\ (\ref{R}),
 which now reads
\beqa
{\cal R} &=& f_N^4(\mu) \frac{1}{4} \psi^4(x_i, \b_i)
\left\{
[ \phi_{123}^2(\mu)+ \phi_{213}^2(\mu) + 4 T_{123} ^2 (\mu) ]^2
\right. \nonumber \\
&\ & \quad
+ \left. [ \phi_{123}^2(\mu) +\phi_{213}^2(\mu) + 4 T_{123} ^2 (\mu) ]
              [ \phi_{132}^2(\mu) +\phi_{312}^2(\mu) + 4 T_{132} ^2 (\mu) ]
 \right\}\, .
\label{Rpsi}
\eeqa

 The parameter $R_{3q}$ is a measure of the transverse size of the
 three-quark Fock state.
 Brodsky, Huang, Lepage and Mackenzie \cite{Banff} have presented arguments on
 how to estimate it. Here, we keep it as a free phenomenological
 parameter and examine the behavior of the p-p amplitude within a range
 of values of $R_{3q} \approx 1/k_{\rm rms}$.
 A crucial point is that $R_{3q} < 1/\Lambda$.
 Since $\psi(x_i,\b_i)$ contains the transverse size information
 we set the maximum of the $b$ integrals to $b_c=1/\Lambda$.

 The only, yet conceptually important, modification in the choice of scales
 $\mu, \, w_m$ is in condition (\ref{lowerbound}),
 where $\mu_{\rm min}$ is replaced by $1/R_{3q}$,
\beq
\mu \, , w_1 \, , w_2 \, , w_3 \geq \frac{1}{R_{3q}} \, ,
\label{Rlowerbound}
\eeq
 since $1/R_{3q}$ is a typical transverse momentum.

 The numerical results for ${\cal A}_{3{\rm s}}$ are shown in figs.\ 7-10 for
the
 various LCDA models and for fixed $\Lambda = 0.20$ GeV. The numbers on the
 $R_{3q}$ curves are the slopes in their respective linear regions.

 Before drawing conclusions from the numerical results, we list
 some of
 the approximations made in the course of this calculation.
  ({\em i}) we have integrated only along the
  direction perpendicular to the scattering plane,
 where the two Lorentz contracted disks intersect. This
 picture becomes less clear-cut as $\theta \rightarrow 0$ and contributions
 from the integration over the azimuthal angle in the impact parameter space
 will have to be considered;
 ({\em ii}) we have assumed that the gaussian, eq.\ (\ref{psi}),
 correctly describes the transverse momentum distribution;
 ({\em iii}) we have neglected all
 hadronic mass scales (proton and constituent quark masses)
 that would normally show up in the transverse wave function $\psi$;
({\em iv}) we have neglected the transverse
 momenta in the hard q-q scatterings $H_m$ \cite{LiS,Li}
 by assuming that their contribution
 in the soft quark region can be captured by taking
 $\alpha_s(w_m^2)= \alpha_s(1/R_{3q}^2)$.
 We suggest, however, that these
 and other shortcomings, such as the neglect of
 higher-order corrections in $U$ and the $H^m$, do not affect the
 main features observed in figs. 7-10 and shared by all LCDA models.

 In figs.~7-10, we see:
 ({\em i}) approximate, but not exact, $t^{-4}$, behavior for the
 amplitude for moderately large $t$, where we find
 ${\cal A}_{3{\rm s}} \sim s/t^{4.2}$.
 We note  the  fit $\d \sigma^{pp}/ \d t \sim t^{-8.4}$ of the FNAL data,
 suggested by the authors of ref. \cite{FNAL}.
 ({\em ii}) A transition from Landshoff
 to dimensional counting behavior,
 whose position is a sensitive function
 of the transverse size $R_{3q}$.
 Since the experiment is decisive about the existence of
 Landshoff behavior at least up to $-t = 15\,{\rm GeV}^2$,
 we can infer that our calculations imply a
 small transverse size for the
 valence state, $R_{3q} \le 0.3$ fm \cite{Banff}.
 A much larger value of $R_{3q}$ would exhibit $t^{-10}$
 behavior in the cross section at lower $t$.

 Our results so far are encouraging, but we recognize that
 the energies and momentum transfers involved are not so
 high that the perturbative picture is applicable without
 reservation.  In particular, we would like to get an idea
 of the importance of soft quarks, and consequently low
 momentum transfers to our results.  When one of the
 scales $\xi_i$ becomes small, the corresponding momentum
 transfer decreases as $\xi_i^2$, and at some point the transverse
 size of the ``hard" scattering may be as large as the
 separation between hard scatterings.

 Thus, as an additional test of the independent scattering method,
 we have recalculated the amplitude with a cutoff $\xi_i> \lambda$
 on soft
 interactions.  Fig.~11 shows the effect of demanding that all
 three of the $\xi_i$ be larger than $\lambda$.  The $\lambda=0.0$
 curve is the reference, and corresponds to the calculations in
 the preceding figures.  The contributions of soft quarks
 are substantial, but not overwhelming.  Demanding that the
 softest quark have 1/20th of each proton's momentum leaves more than
 half the amplitude, while demanding 1/10th reduces the amplitude
 by a factor of about three.  We should keep in mind, however, that
 at collider energies, even a quark with $\xi=0.05$ is quite energetic.
 Of course, at such scales, the quark-quark momentum transfer is
 no longer very large, and a treatment that includes the transverse
 momenta of the wave functions as part of the hard scattering
 \cite{LiS} is probably necessary.  Fig.~12
 shows the same calculation,
 but now
 with the restriction imposed only on the two more energetic
 quarks.  Here we note that most of the time the middle quark has
 at least one-tenth of the energy, and it has more than 1/5th
 about one-third
 of the time.  This shows, we believe, that the independent
 scattering picture, in which there are more than one, and
 sometimes three, independent and well-localized
 scatterings, describes the basic picture of proton-proton
 elastic scattering near the forward direction.

\section{Conclusions}

We have described proton-proton near-forward elastic scattering
with an improved pQCD factorization formalism,
 which describes both short-distance $t^{-10}$ and
 Landshoff $t^{-8}$
 power behavior for moderate momentum transfers with $s \gg -t$.
 The presence of $t^{-8}$ behavior in the cross section implies the
 relevance of
  a new scale, of the order of the average transverse momentum of
 the constituent quarks. The inclusion of a nonperturbative,
 or intrinsic,
 transverse size somewhat smaller than $1/\Lambda$
 seems to be required for
 the description valence state.  Indeed, were the proton
 three-quark
 much larger than $1/\Lambda$, Sudakov effects would produce
 $t^{-10}$ behavior in the kinematic region where
 $t^{-8}$ has been observed.  All of these results require color singlet
 exchange for the independent quark-quark hard scatterings.

 We note that it is most natural for us to compare the
calculated amplitude to  experiment, rather
than the cross section directly, since in elastic scattering,
it is the amplitude and not the cross section that is the
object of computation.  Of course, the range of uncertainty in
the cross section is much larger than in the amplitude.
Nevertheless, it is quite striking that for $|\rho|=1$ in eq.~(\ref{rhointro})
 the normalization of the amplitude works so well, considering the high
 powers involved; see eqs.\ (\ref{fN}), (\ref{N}).

  The transition from Landshoff to short-distance power behavior
 is strongly correlated with the transverse size of the three-quark state
 modeling the proton. We believe that further understanding and technical
 improvements on the theoretical level combined with  measurements
 of the position and the width of the
 transition region could enable us to determine the average
 transverse size and estimate the form of the
 transverse structure from the experimental data.  Evidently,
 near-forward elastic scattering involves
 spatially-separated hard scatterings of colored
 objects.  It is clearly of interest to investigate a
 role for such phenomena in diffractive
 jet and other semi-inclusive cross sections.

 Despite the numerical significance of soft-quark contributions,
 the model described above
 displays a satisfying self-consistency.  In fact,
 we know of no other model for p - p elastic scattering that
 describes the $t^{-8}$ behavior of the elastic scattering cross
 section.  We also think it more than ever of interest to determine
 the normalization and $s$ dependence of the singlet scattering
 amplitude.

\hspace{2cm}

 {\em Acknowledgements \/}
 We are grateful for helpful conversations
 with, and insightful comments by, our colleagues
 Jim Botts,
 Stan Brodsky, John Collins, Claudio Coriano, Peter Landshoff,
 Genya Levin, Hsiang-nan Li, Gregory Korchemsky,
 Al Mueller, Anatoly Radyushkin, John Ralston and Mark Strikman.
 This work was supported in part by the National
 Science Foundation under grant PHY 9309888 and by the Texas National
 Research Laboratory.
\newpage

\appendix

\section{Matrix wave functions}

As an alternative to eq.~(\ref{helspinors}), $Y_{\alpha \beta \gamma}$
 may be expanded in terms of
 spin matrix tensors  $E^{(d)}$, $d=1,2,3$ \cite{HKM},
\beq
Y_{\alpha\beta\gamma} =  \sum_{d=1}^{3} E^{(d)}_{\alpha \beta \gamma}(P,h)
     Y^{(d)}(k_{u_1},k_{u_2},k_d;\mu)\, ,
\label{tensorYE}
\eeq
 for incoming protons ($i=1,2$) and in terms of a conjugate basis
\beq
\overline{Y}_{\alpha \beta \gamma} =  \sum_{d=1}^{3}
\overline{E}^{(d)}_{\alpha \beta \gamma}(P,h)
Y^{(d)}(k_{u_1},k_{u_2},k_d;\mu)\, ,
\label{tensorYbar}
\eeq
 for outgoing protons ($i=3,4$).  The $E$'s and ${\bar E}$'s are given by
\beqa
&\ &E^{(1)}_{\alpha \beta \gamma} =
     E^{-1/2}  (\vsl C)_{\alpha\beta}(\gamma_5 N )_\gamma  \, ,
\hspace{1.5cm}
\overline{E}^{(1)}_{\alpha \beta \gamma} =
 E^{-1/2} (C^{-1} \vsl )_{\alpha \beta}(\overline{N} \gamma_5)_\gamma \, ,
\nonumber \\
&\ &E^{(2)}_{\alpha \beta \gamma} =
   E^{-1/2}   (\vsl \gamma_5 C)_{\alpha\beta}N_\gamma \, ,
\hspace{1.9cm}
\overline{E}^{(2)}_{\alpha \beta \gamma} =
- E^{-1/2}(C^{-1}\gamma_5 \vsl )_{\alpha \beta}\overline{N}_\gamma \, ,
\\
&\ &E^{(3)}_{\alpha \beta \gamma} =
i E^{-1/2} (\sigma_{\mu \nu} v^\nu C)_{\alpha\beta}
(\gamma^\mu \gamma_5 N)_\gamma \, ,
\hspace{.25cm}
\overline{E}^{(3)}_{\alpha \beta \gamma} =
-i E^{-1/2} (C^{-1} \sigma_{\mu \nu} v^\nu)_{\alpha \beta}
(\overline{N} \gamma_5 \gamma^\mu)_\gamma \, .
\nonumber
\label{E}
\eeqa
 We have defined these spin structures to be dimensionless.
 Here, $C$ is the charge conjugation matrix, while
\beq
\sigma_{\mu \nu}=\frac{i}{2} [\gamma_\mu,\gamma_\nu],
\hspace{1cm}
\gamma_5= \frac{i}{4!} \epsilon_{\mu \nu \kappa \lambda}
\gamma^{\mu}\gamma^{\nu}\gamma^{\kappa} \gamma^{\lambda} \, ,
\label{gamma5}
\eeq
with $\epsilon_{0123}=1$ and $N(P,h)$ the massless proton spinor,
normalized, as above, to
\beq
\overline{N}(P,h) \gamma^\mu N(P,h) = 2 P^\mu \, .
\label{spinorm2}
\eeq

The real functions $Y^{(d)}$ in this basis
are conveniently normalized in momentum space as
\beq
Y^{(d)}(k_{u_1},k_{u_2},k_{d};\mu) = \frac{1}{ 2^{1/4} 4 N_{c}!}
f_{N}(\mu)
\Psi^{(d)}(k_{u_1},k_{u_2},k_{d};\mu) \, ,
\label{FourierY}
\eeq
where, in a variation of the  notation of ref.~\cite{CZnuclphys}
for instance,
\beq
\Psi^{(1)} \equiv V^\prime,
\hspace{1cm}
\Psi^{(2)} \equiv A^\prime,
\hspace{1cm}
\Psi^{(3)} \equiv T^\prime \, ,
\label{VAT}
\eeq
  $f_{N}(\mu)$ is a
  standard normalization constant, specified below.
 The primes indicate that these wave functions still depend
 on all four components $k_i^\mu$ and have non-zero mass dimension.
 The various symmetries of the wave function require that
 $V^\prime, A^\prime$ and $T^\prime$ may be themselves obtained from
 a the single function $\Psi(k_{u_1},k_{u_2},k_{d};\mu)$, as
\beqa
V^\prime(k_{u_1},k_{u_2},k_{d};\mu)
                          &=& 1/2[\Psi(k_{u_1},k_{u_2},k_{d};\mu)+
                          \Psi(k_{u_2},k_{u_1},k_{d};\mu)] \, ,\nonumber \\
A^\prime(k_{u_1},k_{u_2},k_{d};\mu)
                          &=& 1/2[\Psi(k_{u_2},k_{u_1},k_{d};\mu)-
                          \Psi(k_{u_1},k_{u_2},k_{d};\mu)] \, ,\nonumber \\
T^\prime(k_{u_1},k_{u_2},k_{d};\mu)
                          &=& 1/2[\Psi(k_{u_1},k_{d},k_{u_2};\mu)+
                          \Psi(k_{u_2},k_{d},k_{u_1};\mu)] \, .
\label{VATrels}
\eeqa
Eq.~(\ref{imprep}) enables us to relate $V'$, $A'$ and $T'$ to
dimensionless
functions of the light-cone variables $\xi_i$ only
(the functions $V$, $A$ and $T$ of Ref.~\cite{CZnuclphys}).

 Linear combinations of the spin structures $E^{(d)}$
 parameterize the on-shell states with definite quark helicity
 defined by eq.~(\ref{helspinors}).
 The correspondence for our
 normalization choice, and for a proton with $(+)$ helicity, is
\beqa
&\ &{\cal M}^{(1)}_{\alpha \beta \gamma} =
 E^{(1)}_{\alpha \beta \gamma}- E^{(2)}_{\alpha \beta \gamma} =
 (E_1 E_2 E_3 /2)^{-1/2} \;
 u_\alpha(k_{u_1},+) \; u_\beta(k_{u_2},-)  \; d_\gamma(k_d,+) \, ,
\nonumber \\
&\ &{\cal M}^{(2)}_{\alpha \beta \gamma} =
 E^{(1)}_{\alpha \beta \gamma}+ E^{(2)}_{\alpha \beta \gamma} =
 (E_1 E_2 E_3 /2)^{-1/2} \;
 u_\alpha(k_{u_1},-) \; u_\beta(k_{u_2},+) \; d_\gamma(k_d,+) \, ,
\nonumber \\
&\ &{\cal M}^{(3)}_{\alpha \beta \gamma} =
 E^{(3)}_{\alpha \beta \gamma} =
- (E_1 E_2 E_3 /2)^{-1/2} \;
 u_\alpha(k_{u_1},+) \; u_\beta(k_{u_2},+) \; d_\gamma(k_d,-)  \, ,
\label{helspinors2}
\eeqa
 where, as in
 Sec.\ 2, $E_1 \; , E_2 \; , E_3$ are the energies of the first and second
 $u$-quarks and the $d$-quark respectively.

  With this form of the wave function, the calculation of the p-p
  amplitude involves Dirac traces
  over the spin structures $E^{(d)}$, which
   are rather lengthy but straighforward. The following relations
 are helpful ($\eta=(+,-,-,-)$):
\beq
( \overline{N}_3 \gamma^\mu N_1) ( \overline{N}_4 \gamma_\mu N_2) =
-2 s \delta_{h_1 h_3} \delta_{h_2 h_4}
\label{spinornorm2}
\eeq
\beqa
i \epsilon_{\mu \nu \rho \sigma} \gamma^\mu \gamma_5  &=&
\gamma_\nu \gamma_\rho  \gamma_\sigma -
\eta_{\nu \rho} \gamma_{\sigma} +
\eta_{\nu \sigma} \gamma_\rho -
\eta_{\rho \sigma} \gamma_\nu  \, ,
\nonumber \\
\frac{1}{16} {\rm tr}
\{ [\gamma_\mu \, , \gamma_\nu] \gamma^\alpha
   [\gamma_\rho \, ,\gamma_\sigma] \gamma^\beta \}  &=&
\eta_{\mu _[\rho} \delta_{\sigma_]}^{_(\alpha} \delta_\nu^{\beta_)} +
\eta_{\nu _[\sigma} \delta_{\rho_]}^{_(\alpha} \delta_\mu^{\beta_)} -
\eta_{\mu _[\rho} \eta_{\sigma_] \nu} \eta^{\alpha \beta} \, .
\label{lemma}
\eeqa
The result, of course, is the same as in eqs.~(\ref{A3P}),
(\ref{T3P}) and (\ref{R}) above.

\newpage

 Table 1.  \newline
 The gluonic exchange channels in the hard scatterings for each flavor flow.

\vspace{0.3cm}

\begin{center}
\begin{tabular}{clc}\hline \hline
$f$  & channels in $H^{1}H^{2}H^{3}$  & no. of channels \\ \hline
1  & $ttt,ttu,tut,tuu$, + $t\leftrightarrow u$      &  8    \\
2  & $ttt,utt$                                      &  2    \\
3  & $tuu,uuu$                                      &  2    \\ \hline \hline
\end{tabular}

\end{center}

\vspace{6cm}

\textwidth=16cm
\oddsidemargin=0.3cm

 Table 2. \newline
 The decomposition coefficients in terms of Appel polynomials,
 and the anomalous dimensions of
 $\phi(x_i;\mu)$, for the models of
 Chernyak and Zhitnitsky (CZ) \cite{CZnuclphys},
 Chernyak, Ogloblin and Zhitnitsky (COZ) \cite{COZ},
 King and Sachrajda (KS) \cite{KS}
 and Gary and Stefanis (GS) \cite{GS}.

\vspace{0.3cm}
\begin{center}
\begin{tabular}{c c l l l l c c}\hline \hline
$j$  & $b_j$ & $a_j$ (CZ) & $a_j$ (COZ) & $a_j$ (KS) & $a_j$ (GS) & $N_j$ &
       $ A_j(x)$  \\ \hline
 0   &  0    & 1.00       & 1.00        & 1.00       & 1.00       &   1   &
        1   \\
 1   & 20/9  & 0.410      & 0.350       & 0.310      & 0.391      &  21/2 &
      $x_1-x_3$ \\
 2   & 24/9  & -0.550     & -0.424      & -0.370     & -0.588     &  7/2  &
      $3x_2-1$ \\
 3   & 32/9  & 0.357      & 0.460       & 0.630      & -0.749     & 63/10 &
      $3(x_1-x_3)^2+x_2(5x_2-3)$ \\
 4   & 40/9  & -0.0122    & -0.00259    & 0.00333    & 0.0176     & 567/2 &
      $ (1/3)(x_1-x_3)(4x_2-1)$ \\
 5   & 42/9  & 0.00106    &  0.0633     &  0.0632    & 0.574      & 81/5  &
      $7x_2-5+(14/3)(x_1^2+x_3^2+3x_1x_3)$ \\ \hline \hline
\end{tabular}

\end{center}

\newpage

 Table 3. \newline
 The slopes of the $b_c$-lines in figs. 3, 4 considered approximately straight
 in the $t$-regions (a), (b), (c).

\vspace{0.3cm}
\begin{center}
\begin{tabular}{c c c c c c c }\hline \hline
	      & \multicolumn{3}{c}{slopes for $\Lambda=0.1$ GeV}
              & \multicolumn{3}{c}{slopes for $\Lambda=0.2$ GeV} \\
$b_c\Lambda$  &  (a)  &  (b)  &  (c)  &  (a)  &  (b)  &  (c)  \\  \hline
  1.0         & -0.38 & -0.61 & -0.84 & -0.23 & -0.52 & -0.89 \\
  0.8         & -0.28 & -0.52 & -0.81 & -0.17 & -0.42 & -0.79 \\
  0.6         & -0.19 & -0.39 & -0.70 & -0.10 & -0.33 & -0.69 \\
  0.4         & -0.09 & -0.27 & -0.56 & -0.02 & -0.23 & -0.58 \\ \hline \hline
\end{tabular}

\end{center}

\newpage

\begin{center}
{\bf Figure Captions}
\end{center}

\begin{enumerate}
\item  Proton-proton elastic scattering in the multiple scattering scenario,
       and flavor routings for each flavor flow $f$. The dashed lines
       represent the $d$-quarks. All momenta flow from left to right.

\item Graphs whose IR divergences mutually cancel.

\item The three-singlet amplitude $|{\cal A}_{3{\rm s}}|$, modulo
      $\rho^3 N s/t^4$, as a
      function of $-t$ for the CZ model
      \cite{CZnuclphys}, without transverse structure and
      for $\Lambda=0.1$ GeV, $\mu_{\rm min}=1.0$ GeV.

\item $|{\cal A}_{3{\rm s}}|$, modulo
      $\rho^3 N s/t^4$, as a
      function of $-t$ for the CZ model
      \cite{CZnuclphys}, without transverse structure and
      for $\Lambda=0.2$ GeV, $\mu_{\rm min}=1.0$ GeV.

\item $|{\cal A}_{3{\rm s}}|$, modulo
      $\rho^3 N s/t^4$, as a
      function of $-t$ for the CZ model
      \cite{CZnuclphys}, without transverse structure and
      for $\Lambda=0.1$ GeV, $\mu_{\rm min}=0.7$ GeV.

\item $|{\cal A}_{3{\rm s}}|$, modulo
      $\rho^3 N s/t^4$, as a
      function of $-t$ for the CZ model
 \cite{CZnuclphys}, without transverse structure and
      for $\Lambda=0.2$ GeV, $\mu_{\rm min}=0.7$ GeV.

\item $|{\cal A}_{3{\rm s}}|$, modulo
      $\rho^3 N s/t^4$, as a
      function of $-t$ for the CZ model
      \cite{CZnuclphys}, with transverse structure and
      for $\Lambda=0.2$ GeV. The ISR data are from ref. \cite{ISR} at
      $\sqrt{s}=52.8$ GeV, and
      the FNAL data from ref. \cite{FNAL} at $\sqrt{s}=27.4$ GeV.
      The error bars include both statistical and normalization errors.

\item $|{\cal A}_{3{\rm s}}|$, modulo
      $\rho^3 N s/t^4$, as a
      function of $-t$ for the COZ model
      \cite{COZ}, with transverse structure and
      for $\Lambda=0.2$ GeV.

\item $|{\cal A}_{3{\rm s}}|$, modulo
      $\rho^3 N s/t^4$, as a
      function of $-t$ for the KS model
      \cite{KS}, with transverse structure and
      for $\Lambda=0.2$ GeV.

\item $|{\cal A}_{3{\rm s}}|$, modulo
      $\rho^3 N s/t^4$, as a
      function of $-t$ for the GS model
      \cite{GS}, with transverse structure and
      for $\Lambda=0.2$ GeV.

\item $|{\cal A}_{3{\rm s}}|$, modulo
      $\rho^3 N s/t^4$, with lower cutoff $\lambda$ on the scale $\xi_{\rm
med}$
      of each of
      the hard scatterings.

\item $|{\cal A}_{3{\rm s}}|$, modulo
      $\rho^3 N s/t^4$, with lower cutoff $\lambda$ on the scale $\xi_{\rm
med}$
      of
      the intermediate hard scattering.

\end{enumerate}

\end{document}